\documentclass[aps,prl,twocolumn,groupedaddress,showpacs]{revtex4-1}
\usepackage[normalem]{ulem}
\usepackage{graphicx}
\usepackage{color}
\usepackage{bbold}
\usepackage{amsmath}
\usepackage{hyperref}

\newcommand{\hs}{\hspace{0.25cm}}
\newcommand{\fullstop}{\hs.}
\newcommand{\comma}{\hs,}

\newcommand{\Rxn}[1][n]{\vec{R}^{(#1)}}
\newcommand{\Rixn}[1][n]{\vec{R}_i^{(#1)}}
\newcommand{\Rbxn}[1][n]{\vec{R}_b^{(#1)}}
\newcommand{\Rdxn}[1][n]{\vec{R}_d^{(#1)}}

\DeclareMathOperator{\Tr}{Tr}

\begin{document}

\title{Failure of the Volume Function in Granular Statistical Mechanics and an Alternative Formulation}
\author{Raphael Blumenfeld$^{1,2,3}$, Shahar Amitai$^1$, Joe F. Jordan$^1$ and Rebecca Hihinashvili$^1$ }

\affiliation{1. Imperial College London, London SW7 2BP, UK\\
2. College of Science, NUDT, Changsha, 410073 Hunan, PRC\\
3. Cavendish Laboratory, Cambridge CB3 0HE, UK}

\date{\today} 

\begin{abstract} 

We first show that the currently accepted statistical mechanics for granular matter is flawed. The reason is that it is based on the volume function, which depends only on a minute fraction of all the structural degrees of freedom and is unaffected by most of the configurational microstates. Consequently, the commonly used partition function underestimates the entropy severely. We then propose a new formulation, replacing the volume function with a {\it connectivity} function that depends on all the structural degrees of freedom and accounts correctly for the entire entropy. We discuss the advantages of the new formalism and derive explicit results for two- and three-dimensional systems. We test the formalism by calculating the entropy of an experimental two-dimensional system, as a function of system size, and showing that it is an extensive variable.

\end{abstract} 


\maketitle

The field of granular physics is in urgent need of equations of state, the traditional provider of which is statistical mechanics (SM). Yet, although a granular statistical mechanical formalism was introduced a quarter of a century ago \cite{EdOa89,MeEd89,Ed90}, no such equations have been derived yet. Granular SM is entropy-based. Part of the entropy is structural \cite{EdOa89,MeEd89,Ed90} and corresponds to the different spatial arrangements of the grains, with each structural configuration regarded as a microstate. These microstates depend on $N_s d$ structural degrees of freedom (DOFs) in $d$ dimensions, with $N_s$ the number of contact position vectors (see below). The volume sub-ensemble is based on a volume function $\mathcal{W}$, which is analogous to the Hamiltonian in thermal SM. Namely, the probability that the system be at a structural microstate with volume $V$ is presumed to be $e^{- V / X_0}$, in analogy to the Boltzmann factor $e^{- E / k_B T}$. The factor $X_0 = \partial \langle \mathcal{W} \rangle / \partial S$, called the compactivity, is the analog of the temperature in thermal SM \cite{EdOa89,MeEd89,Ed90}. Every grain configuration can support an ensemble of different boundary forces, each giving rise to a different internal stress microstate \cite{EdBl05,BlEd06,Heetal07,Puetal10,Bletal12,Bietal13}. The boundary forces, $\vec{g}_m\ (m = 1, ..., M)$ are the DOFs that determine the stress microstates. The combined partition function is
\begin{equation} \label{CombPartFun0}
Z = \int e^{- \frac{\mathcal{W}\left(\{\vec{r}\}\right)}{X_0} - \sum_{ij}\frac{\mathcal{F}_{ij}\left(\{\vec{r}\},\{\vec{g}\}\right)}{X_{ij}}} \prod_{n = 1}^{N_s} d \vec{r}_n\prod_{m = 1}^{M} d \vec{g}_{m} \comma
\end{equation}
where $\sigma_{ij}$ is the stress tensor, $\mathcal{F}_{ij} = V \sigma_{ij}$ is the force moment tensor, and $X_{ij} = \partial \langle \mathcal{F}_{ij} \rangle / \partial S$ is the angoricity tensor \cite{EdBl05,Bletal12}. The identity of the structural DOFs, $\vec{r}$, is discussed below. The two sub-ensembles are not independent \cite{Bletal12} and the total entropy, $S$, is the logarithm of the total number of microstates, both structural and stress. Numerical and experimental tests of the formalism abound \cite{Bietal15,Noetal98, Scetal05, Mcetal09, Soetal12, AsDi08} and some inconsistencies were observed \cite{SoSc14}. In particular, that the compactivity does not equilibrate in some systems \cite{PuDa13}.

Here, we first show that this stems from a fundamental problem with the formulation of the volume ensemble - the volume function, $\mathcal{W}$ in (\ref{CombPartFun0}), is flawed in that it is independent of most of the structural microstates that it is supposed to describe. Consequently, it fails to account correctly for the entire entropy. We then propose an improved formulation that both accounts for all the microstates and is amenable to analytic treatment. We use the new formulation to calculate the new partition function and the mean volume in two ($2d$) and three dimensions ($3d$). The mean volume calculation supports a recent claim that an equipartition principle exists in these systems \cite{Bletal12,AlLu02,WaMe08}. The problem with the volume function is independent of the magnitudes of the boundary forces $\vec{g}_m$. Therefore, for clarity, we take these to be negligibly small, which allows us to neglect the force dependent term in (\ref{CombPartFun0}). Including large boundary forces is straightforward but irrelevant for our purpose here.

In thermal systems, the partition function is a sum over all microstates, each involving a unique combination of the values of the DOFs, giving rise to a specific value of the Hamiltonian, $H$, and hence of the Boltzmann factor. Therefore, $H$ must depend on {\it all} the DOFs. If its derivative with respect to any DOF vanishes identically then $H$ is an incorrect measure of the system's energy and it leads to miscounting of the microstates and miscalculation of the entropy. Thus, dependence on all the DOFs is an essential test of any Hamiltonian-replacing function in granular SM. We demonstrate below that the volume function not only fails this test but it is also independent of {\it almost all} the structural DOFs!

\begin{figure}[h]
\includegraphics[width=0.4\textwidth]{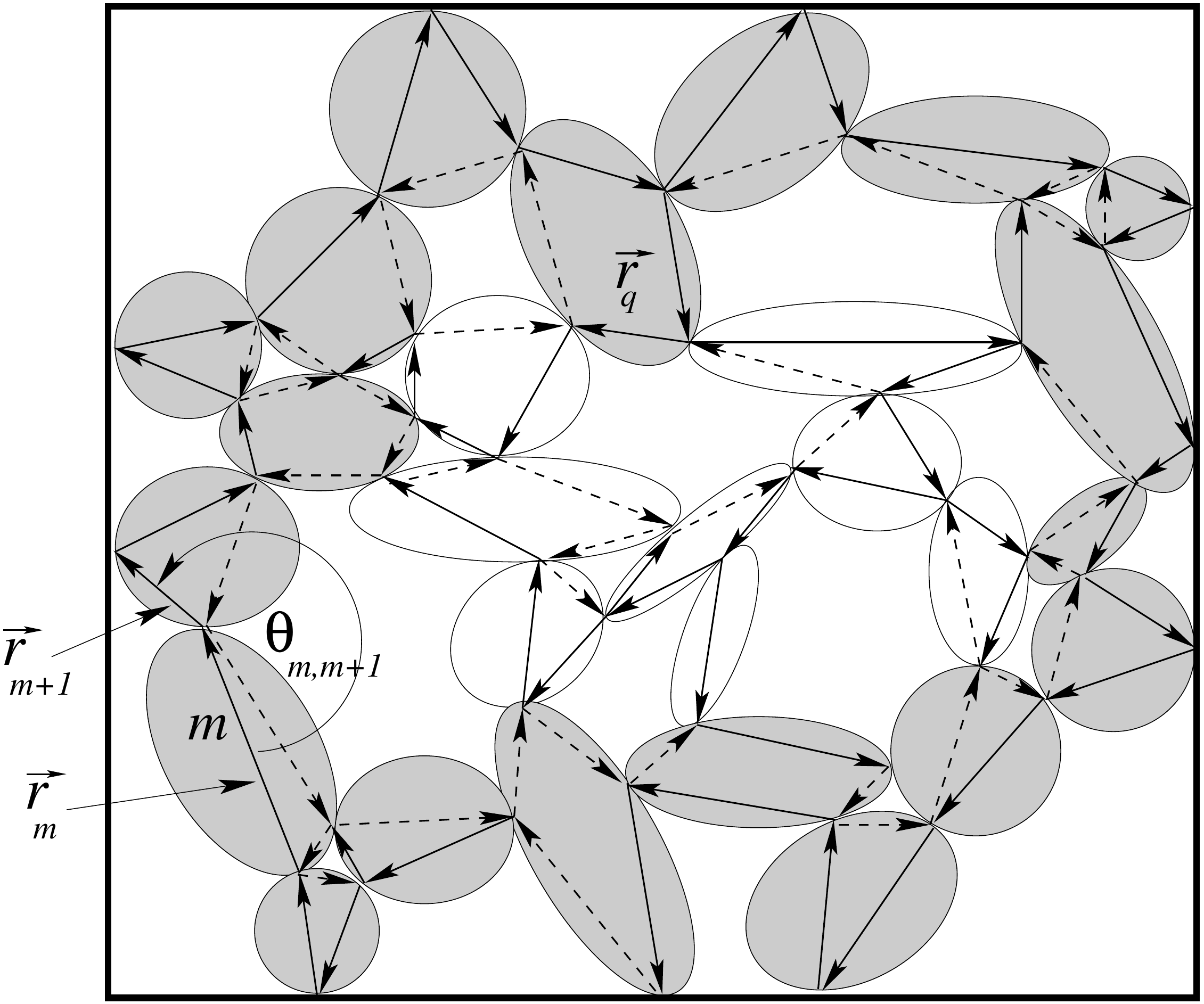}
\caption{A $2d$ granular pack, with $(\alpha - 1) M = 19$ boundary grains (shaded), of which $M=10$ contact the walls. The (solid and dashed) vectors $\vec{r}$ connect a grain's nearest contacts, circulating clockwise. In $3d$, the inter-contact vectors circulate clockwise around the facet that faces a cell, as seen from inside the grain. The solid vectors in the figure form a non-directional spanning tree: they are independent, representing the independent DOFs, they reach every contact and the dashed vectors are linear combinations of them. There are $\alpha M = 29$ boundary vectors, and our choice of spanning tree includes all of them but one.}
\label{GenPack2D}
\end{figure}

We consider an ensemble of identically-generated static systems in $d$ dimensions, comprising all the mechanically equilibrated configurations constructed from a collection of $N \gg 1$ grains. For simplicity, we constrain the mean coordination number, $\bar{z}$, to be the same for each system in the ensemble. Let $M \sim \sqrt{N} \ll N$ and $M \sim N^{2 / 3} \ll N$ be the number of grains in contact with the boundary walls in $2d$ and $3d$, respectively. The total number of boundary grains is $(\alpha - 1) M > M$ ($\alpha = O(1))$ and includes grains that do not touch the walls (see fig. \ref{GenPack2D}). The connectivity is determined uniquely by the intergranular contact position vectors and it is convenient to parameterize these by the vectors, $\vec{r}$, connecting nearest contacts around grains \cite{BlEd03,BlEd06}. In $2d$, these vectors run clockwise around each grain (fig. \ref{GenPack2D}) and a similar parameterization exists in $3d$ \cite{BlEd06,Fretal08}. In both $2d$ and $3d$, there are $(N \bar{z} - M) / 2$ internal intergranular contacts and $\left( N \bar{z} + M \right) / 2$ contacts altogether.

\begin{figure}[h]
\includegraphics[width=0.4\textwidth]{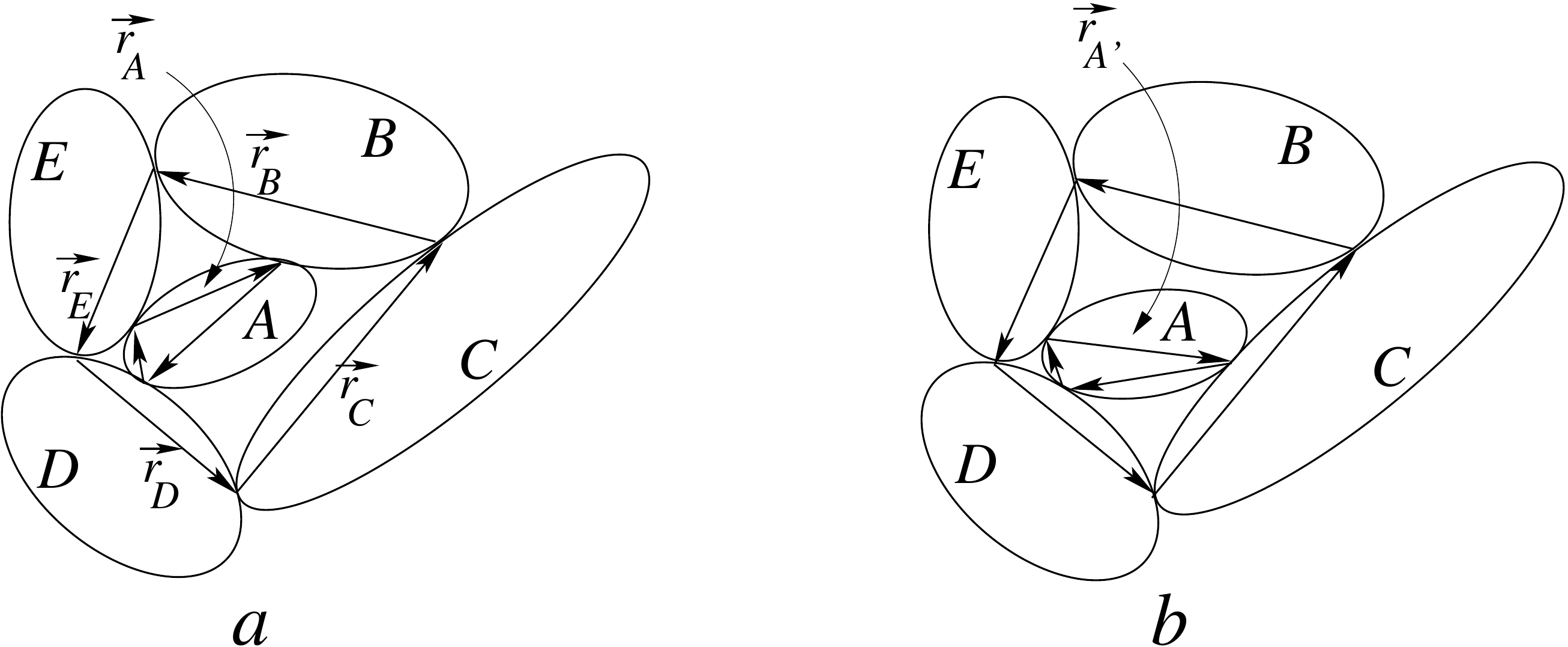}
\caption{The volume function (\ref{Example2DVolume}) does not depend on the inter-contact vectors $\vec{r}$ surrounding grain $A$ and therefore cannot distinguish between configurations $a$ and $b$.}
\label{FlipExample2D}
\end{figure} 
To illustrate the problem with the volume function consider the the example in fig. \ref{FlipExample2D}a. Its volume is
\begin{equation} \label{Example2DVolume}
\mathcal{W} = \frac{1}{2} \big( \mid \vec{r}_B \times \vec{r}_C \mid + \mid \left(\vec{r}_B + \vec{r}_C \right) \times \vec{r}_D \mid \big) \fullstop
\end{equation}
We neglect the contours of the boundary grains extending outside the boundary vectors $\vec{r}_B$-$\vec{r}_E$, whose relative contribution to the total volume is negligible for $N \to \infty$. The key point is that $\mathcal{W}$ does not depend on the inter-contact vectors surrounding grain $A$. Shifting grain $A$ as in fig. \ref{FlipExample2D}b, the volume of the system is still described by (\ref{Example2DVolume}), which depends only on the unchanged DOFs. Thus, $\partial W / \partial \vec{r}_A = \partial W / \partial \vec{r}_{A'} \equiv 0$ and $\mathcal{W}$ cannot register that the configurations in figs. \ref{FlipExample2D}a and \ref{FlipExample2D}b are different. 

This argument is general - the volume function of any $2d$ pack is (see fig. \ref{GenPack2D})
\begin{equation} \label{GenPack2DVolume}
\mathcal{W} = \frac{1}{2} \sum_{m = 1}^{\alpha M - 2} \left| \sum_{k = 1}^{m} \vec{r}_k \times \vec{r}_{m + 1} \right| \comma
\end{equation}
where $\vec{r}_m$ extends between two nearest contacts of boundary grain $m$, $1 \leq m \leq \alpha M$. This function depends only on the boundary contacts - it is independent of {\it any} of the interior ones. For $N \to \infty$, the boundary length scales as $\sqrt{N}$ while the number of configurations can be estimated as $N! \sim N^N$. In contrast, the boundary grains can make order $\sqrt{N}! \sim {\sqrt{N}}^{\sqrt{N}}$ configurations. Thus, the volume function can register only $\sqrt{N}^{\sqrt{N}} / N^N = N^{\sqrt{N} / 2 - N}$ of all the pack's configurations - a minute fraction! This is equivalent to describing a gas in a room by a Hamiltonian that depends only on the gas molecules closest to the walls. Clearly, such a Hamiltonian cannot account for all the entropy of the system. Similarly, the volume function cannot be a good descriptor of the granular entropy. Similarly, volume functions in $3d$ depend only on the boundary inter-contact vectors, i.e. only on $N^{\left( 2N^{2 / 3} / 3 - N \right)}$ of the total number of configurations - a vanishingly small fraction.

Note that this problem with the volume function is more basic than the recently reported failure of the uniform measure assumption in certain systems \cite{Asetal14}, which can be overcome by introducing a non-uniform measure in (\ref{CombPartFun0}). Our new formulation below is similarly independent of this issue and can be used with any measure.

Having concluded that the volume function is not a good equivalent of the Hamiltonian, the question is what could replace it. We propose a connectivity function, $\mathcal{C}$, that does not suffer from these limitations
\begin{equation} \label{CFun}
\mathcal{C} = \sum_{q, p = 1}^{N \bar{z}} \sum_{\alpha, \beta = 1}^d b_{q p ; \alpha \beta} r_{q \alpha} r_{p \beta} \comma
\end{equation}
where the sum is over all the $\vec{r}$-vectors in the system. The coefficients $b_{q p ; \alpha \beta}$ will be identified below. The term $\mathcal{W} / {X_0}$ in (\ref{CombPartFun0}) is then replaced by $\mathcal{C} / \tau$. We name
\begin{equation} \label{QDef}
\tau = \partial \langle \mathcal{C} \rangle / \partial S \comma
\end{equation}
the `contacture' and it replaces the compactivity $X_0$. Here $S$ is the entropy, i.e. the logarithm of the number of all the possible configurations that the packing can be arranged into under the ensemble's constraints. $\tau$ is a measure of the connectivity fluctuations - its increase corresponds to more porous and less compact structures.

To determine the coefficients $b_{q p ; \alpha \beta}$, we require that $\mathcal{C}$ be additive, i.e. that the entropy of a system, made up of two subsystems, is the sum of their entropies. This constrains $b_{q p ; \alpha \beta}$ to have no cross terms and to be independent of $q$ and $p$. Additivity also constrains the connectivity function to be a sum over {\it all} the $\vec{r}$-vectors, rather than only over an independent subset of them. We also require that $\mathcal{C}$ be independent of the coordinate system orientation, constraining $b_{q p ; \alpha \beta}$ to be a scalar constant times the unit matrix. The constant can be absorbed into the definition of $\tau$ and we have
\begin{equation} \label{CFun2}
\mathcal{C} = \sum_{q = 1}^{N\bar{z}} \vec{r}_q \cdot \vec{r}_q = \sum_{n = 1}^d \Rxn \cdot \Rxn \comma
\end{equation}
where $\Rxn \equiv \left(r_{1x_n}, r_{2x_n}, ... \right)$ is a vector of the $x_n$ component of all the $\vec{r}$-vectors. This connectivity-based formulation is not only sensitive to all the structural microstates, but $\mathcal{C}$ also has two significant advantages: it has the same units in all dimensions, as the energy in conventional SM and unlike the volume function, and it is quadratic, enabling analytic calculations of the partition function, as shown below. 

Expression (\ref{CFun2}) is not as innocuous as it may look since only $N_s - 1$ of all the $\vec{r}$-vectors are independent. We separate $\Rxn$ into three sub-vectors, $\Rxn = ( \Rixn, \Rbxn, \Rdxn )$: $\Rixn$ contains the $x_n$ component of the internal independent vectors and is $(N_s - \alpha M)$-long (see below); $\Rbxn$ contains the independent boundary contact vectors, of which there are $\alpha M - 1$ (see below); and $\Rdxn$ contains all the remaining $N_d$ dependent vectors, which can be expressed in terms of $\Rixn$ and $\Rbxn$ as: $\Rdxn = A_1 \cdot \Rixn + A_2 \cdot \Rbxn$, where $A_1$ and $A_2$ are, respectively, $N_d \times \left(N_s - \alpha M \right)$ and $N_d \times \left( \alpha M - 1 \right)$ matrices. In terms of the independent vectors, we have
\begin{align} \label{CFun3}
\begin{split}
\mathcal{C} &= \sum_{n = 1}^d \bigg[ \Rixn \cdot \Rixn + \Rbxn \cdot \Rbxn \\
&+ \left( A_1 \Rixn + A_2 \Rbxn \right) \cdot \left( A_1 \Rixn + A_2 \Rbxn \right) \bigg] \fullstop
\end{split}
\end{align}
The independent $\vec{r}$-vectors, of which there are many choices, form a (non-directional) spanning tree on the contact network. We constrain our choice to include the $\alpha M - 1$ independent boundary contact vectors (fig. \ref{GenPack2D}), as this makes it easier to calculate the partition function. Interestingly, this number holds both in $2d$ and in $3d$, which is shown as follows. In $2d$, the boundary is a closed perimeter of $\alpha M$ vectors, of which $\alpha M - 1$ are clearly independent. In $3d$, the boundary is a $2d$ surface, made of $\alpha M$ nodes and $\zeta \alpha M / 2$ vectors, where $\zeta$ is the surface's mean number of contacts per grain. Using Euler topological relation for planar graphs, this surface consists of $\left( \zeta / 2 - 1 \right) \alpha M - 1$ elementary loops, each of which has one dependent $\vec{r}$. Thus, in $3d$, there are $\zeta \alpha M / 2 - \left[ \left( \zeta / 2 - 1 \right)\alpha M - 1 \right] = \alpha M - 1$ independent surface vectors. 

Using (\ref{CFun3}), the connectivity partition function is
\begin{align} \label{CombPartFun1}
Z = &\int e^{ - \sum_{n=1}^d \left( \Rixn \cdot B_1 \cdot \Rixn + \Rbxn \cdot B_2 \cdot \Rbxn + \Rbxn \cdot B_3 \cdot \Rixn \right)/\tau} \nonumber \\
&\times \prod_{n=1}^d d^{N_s - \alpha M} \Rixn d^{\alpha M - 1} \Rbxn \comma
\end{align}
where $B_1, B_2, B_3 = \mathbb{1} + A_1^T \cdot A_1, \mathbb{1} + A_2^T \cdot A_2, 2 A_2^T \cdot A_1$, respectively. The exponential makes the contribution of large $\vec{r}$-vectors to the partition function negligibly small, allowing us to extend the integration to $\infty$. The contribution of very small $\vec{r}$-vectors is also negligible, allowing us to ignore their absence. Integrating first over $\Rixn$ and then over $\Rbxn$ gives the structure partition function
\begin{equation} \label{CombPartFun2}
Z = \left( \frac{ \left( \pi \tau \right)^{N_s - 1}} {\mid B_1 \mid \mid E \mid} \right)^{d/2} \comma
\end{equation}
where $E \equiv B_2 - \frac{1}{4} B_3 \cdot B_1^{-1} \cdot B_3^T$. 
The mean connectivity is $\langle C \rangle = \tau^2 \partial_\tau \ln Z = (N_s - 1) d\tau/2$. We see that $\langle C \rangle$ is shared amongst all the DOFs, establishing a granular equipartition principle similar to the one obtained in \cite{Bletal12}. Explicitly, the entropy is:
\begin{equation} \label{CombPartFun3}
S = \frac{\langle C \rangle}{\tau} + \ln Z = \frac{d}{2} \Big[ (N_s \! - \! 1) \ln (e \pi \tau) - \ln (|B_1||E|) \Big]
\end{equation}

\begin{figure}[h]
\includegraphics[width=0.3\textwidth]{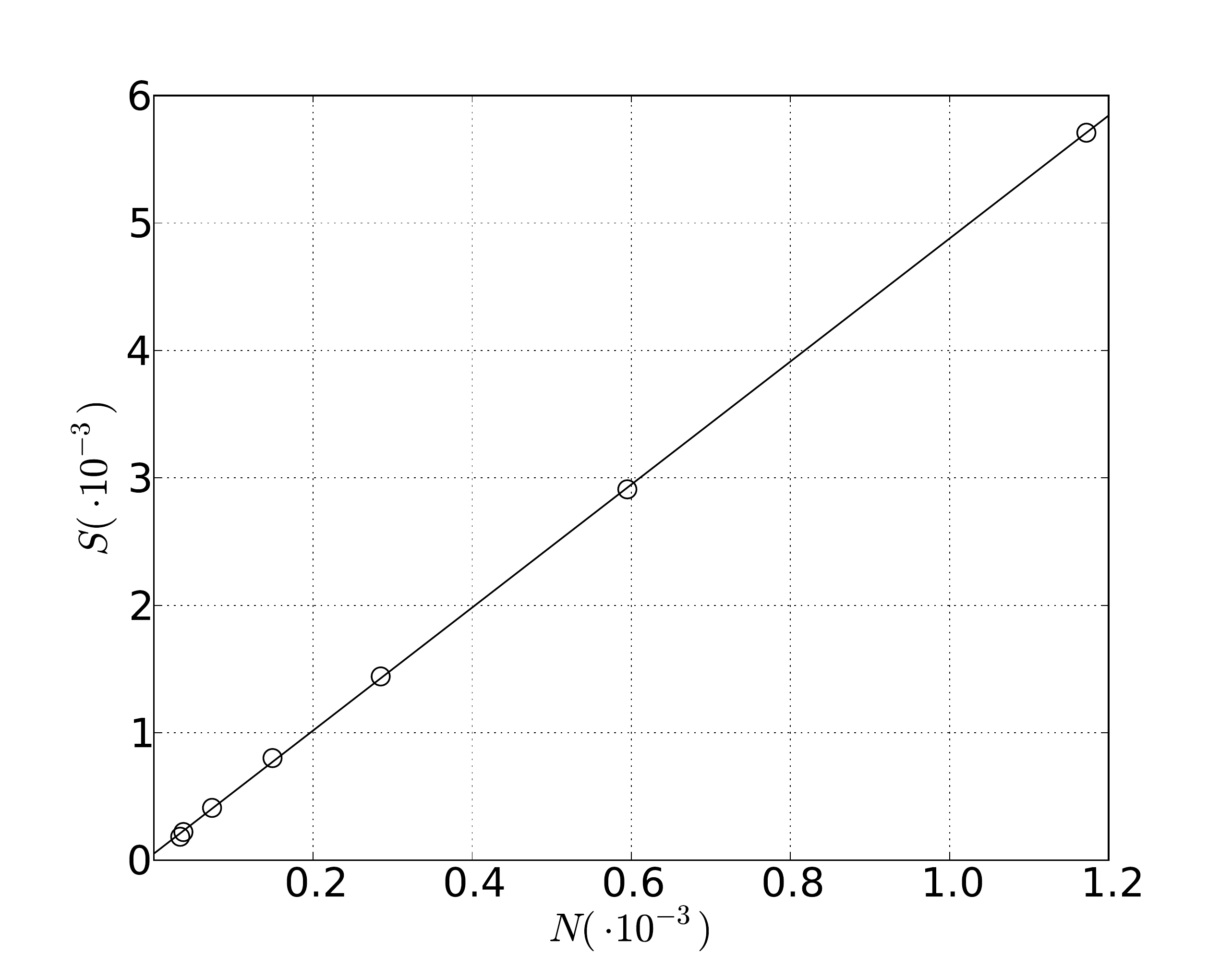}
\caption{Entropy vs. number of particles of the $2d$ experimental granular systems of \cite{Caletal97}).}
\label{EntropyBySize}
\end{figure}

To demonstrate the use of our formalism, we analyse $2d$ experimental systems, each of $1172$ discs of three different radii, produced by the 3SR Lab \cite{Caletal97}. For each system we construct the contact network, choose a spanning tree, express the dependent $\vec{r}$-vectors in terms of the independent ones, calculate the matrices $A_i$, $B_i$ and $E$, and then compute the entropy using eq. (\ref{CombPartFun3}). Fig \ref{EntropyBySize} shows the entropy for non-overlapping subsystems of different sizes. We find that the entropy increases linearly with system size, i.e. it is extensive. The increase rate depends on the unknown $\tau$ (see eq. (\ref{CombPartFun3})). Our relation is linear without subtracting $\ln(N!)$, in contrast to \cite{Asetal14}.

Rewriting (\ref{CombPartFun1}) as $Z = \mathcal{L}(1)$, any structural expectation value is $\langle A \rangle = \mathcal{L}(A) / Z$. E.g. the boundary vectors satisfy
\begin{equation} \label{Rb2}
\langle \vec{r}_{b,q} \cdot \vec{r}_{b,q} \rangle = \frac{\mathcal{L} \left( \vec{r}_{b,q} \cdot \vec{r}_{b,q} \right)}{Z} = \frac{\tau d}{2} \left( E^{-1} \right)_{q, q} \fullstop
\end{equation}
Since $\langle \vec{r}_{b,q}^2 \rangle$ is independent of system size, $\tau$ is inversely proportional to a typical single entry of $E^{-1}$. Similarly, the mean magnitude square of an internal vector is $\langle \vec{r}_{i,q} \cdot \vec{r}_{i,q} \rangle = \tau d \left( G^{-1} \right)_{q,q} / 2$, with $G \equiv B_1 - \frac{1}{4} B_3^T \cdot B_2^{-1} \cdot B_3$. Significantly, we verified numerically that the entries of $E^{-1}$ and $G^{-1}$ are independent of system size, which establishes that $\tau$ is an {\it intensive variable}.

To calculate the mean volume, we use (\ref{GenPack2DVolume}) and define the interior angle between neighbour vectors $q$ and $q + 1$ along the boundary (fig. \ref{GenPack2D}) as $\left(1 - \frac{2}{\alpha M} \right) \pi + \delta \theta_{q, q + 1}$, where $\delta \theta_{q, q + 1}$ is its deviation from that of a regular $\alpha M$-sided polygon. It is straightforward to show that the angle between boundary vectors $\vec{r}_k$ and $\vec{r}_{m + 1}$ is $\sum_{q = k}^{m} \left( \frac{2 \pi}{\alpha M} - \delta \theta_{q, q + 1} \right)$. The $2d$ volume is then
\begin{equation} \label{ExpecV1}
V_{2d} = \frac{1}{2} \sum_{m = 1}^{\alpha M - 2} \sum_{k = 1}^{m} r_k r_{m + 1} \sin \sum_{q = k}^{m} \left( \frac{2 \pi}{\alpha M} - \delta \theta_{q, q + 1} \right) \fullstop
\end{equation}
For $M \! \gg \! 1$, the sum over the constant term $2 \pi / \alpha M$ dominates over the fluctuations $\delta \theta_{q, q + 1}$ and we take it out of the integral. Since $k \neq m + 1$, the integration over $r_k r_{m + 1}$ yields $\langle \mid \! r_b \! \mid \rangle^2 = \langle r_b^2 \rangle$ and, using (\ref{Rb2}), we obtain 
\begin{equation} \label{ExpecV2}
\langle V_{2d} \rangle \approx \frac{\alpha^2 M^2}{2 \pi} \langle r_b^2 \rangle \approx \frac{\alpha^2 M^2 \tau}{2 \pi} U_E \sim N \tau \comma
\end{equation}
where $U_E \equiv \Tr \left\{ E^{-1} \right\} / (\alpha M - 1)$ is the average of the diagonal element in the matrix $E^{-1}$. Since $\tau = O(1)$ then $ \langle V_{2d} \rangle \sim M^2 \sim N_s$ and the mean volume is also shared equally amongst the DOFs - reaffirming the granular equipartition principle \cite{Bletal12}.

In $3d$ we specialise to star-like systems, where all the boundary contact positions are uniquely defined in terms of the angles from an origin in the system. The volume is then a sum over tetrahedra, whose apexes are at one of the system's internal contacts (e.g. the closest to the contact network centroid) and whose bases are the triangular facets that make the network's boundary
\begin{align} \label{V3}
\begin{split}
V_{3d} &= \frac{1}{3} \sum_{n = 1}^{N_{\rm triangles}} \mid \! \left( \vec{r}_{n1} \times \vec{r}_{n2} \right) \cdot \vec{\rho}_{n} \! \mid = \\
&= \frac{1}{3} \sum_{n=1}^{N_{\rm triangles}} \mid \! \left( \vec{r}_{n1} \times \vec{r}_{n2} \right) \cdot \left( \sum_{k = 1}^{K_{n}} \vec{r}_{n k} \right) \! \mid \fullstop
\end{split}
\end{align}
The first sum is over the boundary triangles, $\vec{r}_{n1}$ and $\vec{r}_{n2}$ are two edges of triangle $n$, and $\vec{\rho}_{n}$ is the vector from the tetrahedron apex to the contact point that these two edges share. The second sum is over the $K_n$ independent contact vectors that make $\vec{\rho}_{n}$. The angles between the triangle edges, $\alpha_{n}$, are distributed around $\pi/3$. The angles that the vectors $\vec{r}_{n k}$ make with $\vec{\rho}_{n}$, $\cos \theta_{n k} = \frac{\vec{r}_{n k} \cdot \vec{\rho}_{n}}{\mid r_{n k} \mid \mid \rho_n \mid}$, are distributed around $\theta = 0$. Evaluating the sum by averaging separately over the angles and the magnitudes of the contact vectors, gives 
\begin{align} \label{ExpecV3}
\begin{split}
\langle V_{3d} \rangle &= \frac{N_{\rm triangles} \bar{K_n}}{2 \pi^2} \langle \mid r_b \mid \rangle^2 \langle \mid r_i \mid \rangle \\
&= \frac{(3/2)^{3/2} N_{\rm triangles} \bar{K_n}}{2 \pi^2} U_E U_G^{1/2} \tau^{3/2} \comma
\end{split}
\end{align}
where $\bar{K_n}$ is the mean number of $\vec{r}$-vectors between the system centroid and the boundary triangles. $U_G \equiv \Tr \left\{ G^{-1} \right\} / (N_s - \alpha M)$ is the average of the diagonal element in the matrix $G^{-1}$. From dimensional considerations, only $N_{\rm triangles} \sim N^{2 / 3}$ and $\bar{K_n} \sim N^{1 / 3}$ depend on $N$ in (\ref{ExpecV3}) and hence $\langle V_{3d} \rangle \sim N$. This substantiates the existence of an equipartition principle in $3d$ too.

To conclude, we have pointed out a flaw in the original entropic formalism of granular SM - the volume function depends only on a minute fraction of the DOFs and is insensitive to most microstates. This results in a significant underestimate of the number of microstates and hence of the entropy. We then proposed to replace the volume function by a connectivity function, which is additive and depends on all the structural DOFs. The compactivity must then be replaced by a new measure - the contacture, $\tau$. The new formulation was used to obtain analytical expressions for the entropy and for several expectation values, as well as to analyse $2d$ experimental systems. We verified that the entropy is extensive, $\tau$ is intensive, and calculated the mean volume in $2d$ and $3d$. The mean volume was shown to be proportional to the number of structural DOFs, supporting an equipartition principle \cite{Bletal12,AlLu02,WaMe08}. The flaw pointed out here probably explains the observations in \cite{PuDa13} that the stress-based angoricity equilibrates in subsystems while the volume-based compactivity does not, casting doubt on the usefulness of the compactivity as a good descriptor of the structural fluctuations.

It is difficult to compare our method with recent attempts at static granular statistical mechanics as a glass-like transition \cite{Chetal14, Raetal15}. These rely on studying the jamming dynamics, using conventional positions and momenta, energy and temperature. The lack of ergodicity forces these into questionable relations between energetic and structural ensembles, e.g. that each energetic state corresponds to one structural configuration \cite{Raetal15}. The analysis of the jamming state focuses on a particular, protocol-dependent state, where description based on force DOFs is complete \cite{Leetal13, Geetal15}. At the same time our SM approach addresses the full phase space, including both structural and force DOFs.

A major advantage of the new formulation is the Gaussian form of the partition function in all dimensions, making possible derivation of exact results, as we demonstrated. In particular, it paves the way to an explicit equation of state relating the means of the volume and the stress. It would be interesting to revisit previous analyses with the new formulation, including the coupling between the structure and stress microstates \cite{Bletal12,Bletal13}, and study the contacture equilibration, as in \cite{PuDa13}. We look forward to numerical and experimental tests of the new formulation.

\begin{acknowledgments}
This work has been funded in part by EPSRC - EP/H051716/1 and two Alan Howard PhD Scholarships.
\end{acknowledgments}

\end{document}